# Comment on "Investigation of the anisotropic distribution of microdosimetric quantities in the vicinity of X-ray-irradiated gold nanoparticles" by Derrien et al. [Radiation Physics and Chemistry 213, 111232 (2023)]

**Hans Rabus**
Physikalisch-Technische Bundesanstalt, Berlin, Germany
**Email**: hans.rabus@ptb.de

In their recent article, Derrien et al. (Derrien et al., 2023) study the anisotropy of microdosimetric quantities for spherical sites of several sizes placed around spherical gold nanoparticles of several diameters irradiated by monoenergetic photons. This comment points out that

(1) the reported single event distributions of specific energy may be biased due to overcounting.
(2) by considering only energy imparted by electrons produced in photon interactions in the nanoparticle, the magnitude of the anisotropy is overestimated by up to orders of magnitude with respect to an irradiation under conditions of secondary particle equilibrium.

## Correctness of the single event distributions

In the approach taken by Derrien et al., an event is defined as a photon interaction occuring in the nanoparticle and energy deposits occuring in the site. Fig. 3 in (Derrien et al., 2023) illustrates the scoring procedure used, which considers multiple spherical sites randomly placed inside the torus enclosing all possible sites at a given radial distance $R$ from the center of the nanoparticle and polar angle $\theta$ with respect to the photon beam direction.

The case illustrated in Fig. 3(d) is supposed to demonstrate the efficiency of the sampling which has some conceptual similarity with the variance reduction technique of particle splitting. Here, Derrien et al. do 'event splitting' by using the same energy transfer points resulting from a photon interaction in the nanoparticle for scoring energy imparted in several sites. While there is no doubt that this method makes the simulations statistically more efficient, it also implies that energy transfer points are sampled multiple times. For instance, the energy deposits $\varepsilon_2$, $\varepsilon_3$ and $\varepsilon_5$ are tallied in the green site as well as in the blue site shown in Fig. 3(d) in (Derrien et al., 2023).

This leads to an overcounting. One has to take into account that by using the three targets in this example, there are three instead of one events. This is not accounted for and leads to bias, presumably favouring low values of energy imparted. In this author's view, this approach attempts to derive an estimate of the single event distribution based on the data from a single primary particle track. This is viable if it is assured that the single-event distribution is normalized to unity as it trivially would be when only one site is used. Therefore, in a case as shown in Fig. 3(d) in (Derrien et al., 2023), the three values of energy imparted need to get a weight of 1/3. In the general case, the weight is the number of sites $n_s^p$ inside the torus registering an event in the primary particle history $p$.

Table 1: Fictive example of events registered in sites located at distance $R$ from the nanoparticle at a polar angle $\theta$. The columns correspond to the primary particle histories $p_k$ producing events with their associated weights $w_{FC}^{p_k}$ correcting the bias from interaction forcing. The other rows correspond to the specific energy bins, and the values given indicate the ratio of the number of targets scoring this value of specific energy to the total number of events registered for the respective primary particle history, listed in the last row. The last column gives the number of entries for the specific energy bin.

|  | $p_1$ | $p_2$ | $p_3$ | $p_4$ | $p_5$ | $p_6$ | $p_7$ | $p_8$ | $p_9$ | $p_{10}$ | $p_{11}$ | $p_{12}$ | $p_{13}$ | $p_{14}$ |  |
|---|---|---|---|---|---|---|---|---|---|---|---|---|---|---|---|
| $w_{FC}^{p_k}$ | $w_1$ | $w_2$ | $w_3$ | $w_4$ | $w_5$ | $w_6$ | $w_7$ | $w_8$ | $w_9$ | $w_{10}$ | $w_{11}$ | $w_{12}$ | $w_{13}$ | $w_{14}$ | $n_j$ |
| $z_1$ |  | 1/5 |  |  |  | 2/4 |  |  |  |  |  |  | 1/6 |  | 4 |
| $z_2$ |  | 1/5 |  | 3/3 |  |  |  |  |  |  | 1/4 |  | 1/6 |  | 6 |
| $z_3$ |  | 1/5 |  |  |  | 1/4 | 1 |  |  | 1/2 |  | 1 | 1/6 |  | 5 |
| $z_4$ |  | 1/5 | 2/3 | 1/3 |  |  |  |  |  |  | 1/4 |  |  |  | 5 |
| $z_5$ |  | 1/5 | 1/3 | 1/3 |  |  |  | 1/2 |  | 1/2 | 1/4 |  | 1/6 |  | 7 |
| $z_6$ | 1 |  |  |  |  | 1/4 |  |  |  |  |  |  | 1/6 |  | 3 |
| $z_7$ |  |  |  | 1/3 |  |  |  | 1/2 | 1 |  | 1/4 |  | 1/6 | 1 | 5 |
| $n_s^{p_k}$ | 1 | 5 | 3 | 3 | 3 | 4 | 1 | 2 | 1 | 2 | 4 | 1 | 6 | 1 |  |

This is illustrated in Table 1, which shows a fictive example detailing the contributions of the particle histories that produce events in sites at a given radial distance $R$ from the nanoparticle at a polar angle $\theta$. The first column shows a primary particle history for which only one of the sites registers an event. The second column is a history where 5 of the sites placed in the torus register an event, each site with its own value of specific energy. The third column could be a case as shown in Fig. 3(a) of (Derrien et al., 2023). Here, two of the sites receive a specific energy in the 4$^{th}$ $z$-bin, while the third one scores a $z$-value in the fifth bin. The fourth column shows three sites registering events in the 4$^{th}$, 5$^{th}$ and 7$^{th}$ $z$-bin. The fifth history produces also events in three sites with $z$-values falling in the 2$^{nd}$ bin.

The value in the numerator of Eq. (1) in (Derrien et al., 2023) for the 1$^{st}$ $z$-bin in this example is $w_2+2w_6+w_{13}$, whereas normalizing with the factors shown in Table 1 gives $w_2/5+w_6/2+w_{13}/6$. Of course, the correction for the number of sites with events per primary particle history also changes the denominator of the single-event frequency distribution as shown in Eq. (1) but it is not evident how this impacts the overall frequency distribution. For unbiased sampling of the single-event distribution of specific energy, Eq. (1) in (Derrien et al., 2023) should therefore be replaced with Eq. (1) given below.

$$f_{1,\text{GNP}}(z_j, R, \theta) = \frac{\sum_i^{n_j(R,\theta)} w_{FC}^{p(i)}/n_s^{p(i)}}{\Delta z_j \sum_k^{n_{\text{bin}}} \sum_i^{n_k(R,\theta)} w_{FC}^{p(i)}/n_s^{p(i)}} \qquad (1)$$

Here, $n_s^{p(i)}$ is the number of sites registering the occurrence of an event in the primary particle history $p(i)$ which contributes the $i^{th}$ occurrence of a specific energy value in the $j^{th}$ $z$-bin. $w_{FC}^{p(i)}$ is the correction factor for photon interaction forcing in the nanoparticle in the primary particle history $p(i)$. It should be noted that this correction factor depends only on the primary particle's start position and not on the location of the site.

**Magnitude of the anisotropy**

In Figure 5, Derrien et al. show the dependence on polar angle of the quantity $\bar{z}_{GNP}^P(R, \theta)$ defined by their Eq. (3). In microdosimetric terms, this quantity is the frequency-mean specific energy which can be expressed as the product of the event frequency and the single-event frequency-mean specific energy. It is well known (Rossi and Zaider, 1996; Lindborg and Waker, 2017) that $\bar{z}_{GNP}^P$ converges to the absorbed dose for small site sizes.

The article of (Derrien et al., 2023) emphasizes several times that they ignore the interactions of photons in water and seem to use the work of (Poignant et al., 2021) as a justification. However, the assertion of (Poignant et al., 2021) that the interactions in water do not impact the microdosimetric distributions applies only to the single-event distributions.

The quantity $\bar{z}_{GNP}^P(R, \theta)$ as per Eq. (4) in (Derrien et al., 2023) applies to a multi-event distribution produced by a photon fluence of one photon per cross-sectional area of the GNP. For the case of a spherical nanoparticle with a radius of 50 nm, to which many of the results shown in the work of (Derrien et al., 2023) relate, the corresponding photon fluence is about $1.3\times10^{10}$ cm$^{-2}$. The corresponding absorbed dose to water under charged particle equilibrium, $D_w$, can be estimated from literature data for the mass-energy absorption coefficients (Hubbell and Seltzer, 2004) of photons in water. The corresponding values are given in the third row of Table 2.

The lower two groups of lines in Table 2 show in the respective first two lines the maximal and minimal values of the dose contribution from the nanoparticle for 200 nm and 1000 nm distance of a site of radius 50 nm from a nanoparticle of 50 nm radius. These values were read from Fig. 5 in (Derrien et al., 2023). The arithmetic mean of the values in the first two lines of a block is shown in the respective third lines and is used as an estimate for the isotropic value $\bar{z}_{GNP}^P|_I$. The respective fourth lines show the values of a correction factor $f_c$ given by Eq. (2).

$$f_c = \frac{\bar{z}_{GNP}^P|_I}{\bar{z}_{GNP}^P|_I + D_w} \qquad (2)$$

The respective last lines gives the resulting corrected value of the mean isotropy index for charged particle equilibrium $\eta_{m,CPE}$ as obtained from the values given in the first and third rows of Table 2 in (Derrien et al., 2023) by multiplying with the correction factors.

These corrected values are below 4% for the 30 keV and 85 keV photons and a site location at 200 nm from the nanoparticle. For all other cases they are well below 1 %. In all cases, the anisotropy is smaller than the values of (Derrien et al., 2023) obtained by considering only the energy deposited by electrons emitted from the nanoparticle after a photon interaction. The largest difference occurs for the case of 150 keV photons and a site at 1000 nm distance where the almost 20 % anisotropy reported by (Derrien et al., 2023) reduces to 0.01%.

Table 2: Upper part: Data from (Hubbell and Seltzer, 2004) for the mass energy absorption coefficient at the photon energies, resulting absorbed dose for a fluence of 1 photon per cross-sectional area of a 50 nm-diameter nanoparticle. Lower part: Maximal and minimal values for the multi-event average specific energy contribution from the gold nanoparticle, $\bar{z}_{GNP}^{P}$, the corresponding arithmetic mean for a 50 nm target at 200 nm and 1000 nm radial distance from the center of the nanoparticale center. The further lines give the correction factor for charged particle equilibrium and the resulting corrected anisotropy coefficient.

| Photon energy | | 30 keV | 85 keV | 150 keV |
|---|---|---|---|---|
| Mass energy absorption coefficient | | 0.16 cm²/g | 0.026 cm²/g | 0.028 cm²/g |
| Absorbed dose $D_w$ for a fluence of $1.3\times10^{10}$ cm$^{-2}$ | | $9.78\times10^{-3}$ Gy | $4.50\times10^{-3}$ Gy | $8.56\times10^{-3}$ Gy |
| $R$ = 200 nm | Maximal $\bar{z}_{GNP}^{P}$ | $3.8\times10^{-3}$ Gy | $1.9\times10^{-3}$ Gy | $1.8\times10^{-4}$ Gy |
| | Maximal $\bar{z}_{GNP}^{P}$ | $2.4\times10^{-3}$ Gy | $1.2\times10^{-3}$ Gy | $1.2\times10^{-4}$ Gy |
| | Mean $\bar{z}_{GNP}^{P}\big|_I$ | $3.1\times10^{-3}$ Gy | $1.6\times10^{-3}$ Gy | $1.5\times10^{-4}$ Gy |
| | Correction factor $f_c$ | 0.24 | 0.26 | 0.017 |
| | Anisotropy coefficient $\eta_{m,CPE}$ | 3.5 % | 3.9 % | 0.27 % |
| $R$ = 1000 nm | Maximal $\bar{z}_{GNP}^{P}$ | $1.5\times10^{-4}$ Gy | $3.0\times10^{-5}$ Gy | $7.7\times10^{-6}$ Gy |
| | Maximal $\bar{z}_{GNP}^{P}$ | $0.8\times10^{-4}$ Gy | $2.6\times10^{-5}$ Gy | $4.3\times10^{-6}$ Gy |
| | Mean $\bar{z}_{GNP}^{P}\big|_I$ | $1.2\times10^{-4}$ Gy | $2.8\times10^{-5}$ Gy | $6.0\times10^{-6}$ Gy |
| | Correction factor $f_c$ | 0.012 | $6.2\times10^{-3}$ | $7.0\times10^{-4}$ |
| | Anisotropy coefficient $\eta_{m,CPE}$ | 0.17 % | 0.02 % | 0.01 % |

In addition to the bias in the magnitude of the anisotropy introduced by neglecting charged particle equilibrium, there is a further source of bias in the simulation code. The probability distribution used in Geant4 for sampling the polar angle of an emitted photoelectron (Geant4 collaboration, 2017) strictly applies only to K-shell photoabsorption (Sauter, 1931a, 1931b; Fano et al., 1959; Gavrila, 1959). For the 30 keV photons, however, only ionization of the L-shells of gold is possible (Cullen et al., 1997). Therefore, the angular distribution of photoelectrons at this photon energy may be affected by an additional bias. It should also be noted that at 85 keV photon energy about 20% of the photoabsorption occurs on shells higher than the K-shell (Cullen et al., 1997).

Finally, it must be noted that the simulation results are for the extreme case of a parallel photon beam that may not be respresentative of the photon radiation field in a realistic irradiation geometry where scattered photons from Compton and Rayleigh interactions cover a large range of directions different from a parallel primary radiation beam. Since the Compton-scattered photons have smaller energy, their photoabsorption cross-section with gold atoms are higher. This leads to a further reduction of the anisotropy.

**Conclusion**

When a spherical nanoparticle is irradiated by a parallel photon beam, there is an anisotropy of the energy imparted in sites around the nanoparticle. Under conditions of secondary particle equilibrium, the magnitude of this anisotropy is significantly smaller than what is reported in (Derrien et al., 2023) for the case that only photon interactions in the nanoparticle are accounted for. While (Derrien et al., 2023) were aware of the limitation of their work due to neglecting secondary electron equilibrium, it seems that they did not expect the predicted mean anisotropy index to be overestimated by a factor between 4 and 2000 (depending on photon energy asnd distance from the nanoparticle). In addition, the single event frequency distributions of specific energy shown by (Derrien et al., 2023) are compromised by a bias due to overcounting.